\documentclass[aps,prb,showpacs,twocolumn]{revtex4}
\usepackage{float,graphicx}
\usepackage{psfig}

\usepackage{dcolumn}

\newcommand{\bsim}{\mbox{\raisebox{-0.1cm}{$\;
\stackrel{\textstyle>}{\sim}\;$}}}
\newcommand{\lsim}{\mbox{\raisebox{-0.1cm}{$\;
\stackrel{\textstyle<}{\sim}\;$}}}

\begin{document}

\title{Poor screening and nonadiabatic superconductivity
in correlated systems}

\author{Lilia Boeri$^1$, Emmanuele Cappelluti$^{2,1}$,
Claudio Grimaldi$^3$ and Luciano Pietronero$^{1,4}$}

\affiliation{$^1$Dipart. di Fisica, Universit\`a ``La Sapienza'',
P.le A. Moro 2, 00185 Roma, and INFM UdR RM1, Italy}

\affiliation{$^2$``Enrico Fermi'' Center,
v. Panisperna 89a, c/o Compendio del Viminale, 00184 Roma, Italy}

\affiliation{$^3$Institut de Production et Robotique, LPM,
\'Ecole Polytechnique F\'ed\'erale de Lausanne,
CH-1015 Lausanne, Switzerland}

\affiliation{$^4$CNR, Istituto di Acustica ``O.M. Corbino'', v. del Fosso
del Cavaliere 100, 00133 Roma, Italy}

\date{\today}

\begin{abstract}
In this paper we investigate the role of the electronic correlation
on the hole doping dependence of electron-phonon and superconducting
properties of cuprates. We introduce a simple analytical expression
for the one-particle Green's function in the presence of electronic
correlation and we evaluate the reduction of the screening properties
as the electronic correlation increases by approaching half-filling.
The poor screening properties play an important role within
the context of the nonadiabatic theory of superconductivity.
We show that a consistent inclusion of the reduced screening properties
in the nonadiabatic theory can account in a natural way for the
$T_c$-$\delta$ phase diagram of cuprates. Experimental evidences
are also discussed.
\end{abstract}
\maketitle

\section{Introduction}

The role of the electron-phonon (el-ph)
interaction in the high-$T_c$ superconducting
cuprates has been a matter of debate for a long time.
In early times the report of a negligible isotope effect on $T_c$
at optimal doping, the almost linear behaviour of the resistivity on
temperature, also at optimal doping, and other exotic features of the
copper oxides led to the common belief that electron-phonon coupling
was a marginal ingredient to understand the phenomenology of these materials.
However, over the years, there has been a revamping evidence
of an important role of the phonons.
The most remarkable ones are, for instance, the discovery
of an isotope effect on $T_c$ larger than the BCS value
($\alpha_{T_c} > 0.5$) in the underdoped regime,\cite{crawford,franck}
the report of a sizable isotope shift on the effective electronic
mass $m^*$\cite{keller} and on the
onset of the pseudogap,\cite{raffa,rubio}
the observation of phonon renormalization\cite{dastuto}
and phonon anomalies at $T < T_c$.\cite{friedl}
More recently, ARPES measurements pointed out a kink in the electron
dispersion the origin of which is probably phononic.\cite{lanzara}
Clearly, if phonons
are relevant for superconductivity in these materials, this
cannot be described in a BCS-like framework, but some non-conventional
approach including strong electronic correlation is necessary.
The study of the interplay between electron-phonon interaction and the
electronic correlation is a challenging task which has attracted
much scientific work along different lines.

An interesting issue concerns the  momentum modulation
of the electron-phonon coupling induced
by the electronic correlation.
In Ref. \onlinecite{kulicrep}, using a variety of
theoretical and experimental findings, it is shown that
in correlated systems small-${\bf q}$ scattering in
the electron-phonon interaction is strongly favored.
A strong enhancement of the forward scattering at ${\bf q} \sim 0$ in
correlated systems close to the
metal-insulator transition, accompanied
by a suppression of scattering at large ${\bf q}$,
was reported for example in
Refs. \onlinecite{zeyher,grilli} by using $1/N$
expansion techniques. A recent numerical work based
on Quantum Monte Carlo technique confirms this picture.\cite{huang}

Different but somehow complementary argumentations
based on poor screening effects in correlated systems
have been also discussed in literature.
The basic idea is that, as a metal
loses its coherence as function of the correlation degree
approaching a metal-insulator transition,
the screening properties of the bare long-range electron-phonon interaction
become less effective resulting in a net predominance of
small ${\bf q}$ scattering.\cite{weger,fay,abrikosov,kotliarcoulomb}
A similar physical argument applies, for example, to
doped semiconductors
which are commonly described in terms of the Fr\"ohlich Hamiltonian,
with electron-phonon matrix elements $|g_{\bf q}|^2 \propto
1/|{\bf q}|^2$.

The momentum structure of the electron-phonon scattering
induced by the electronic correlation has been shown to play
a crucial role in the context of nonadiabatic
superconductivity.\cite{PSG,GPS,GPSprl}
In narrow band systems, such as cuprates and fullerenes, the Fermi energy
$E_{\rm F}$ is so small to be comparable with the phonon frequencies
$\omega_{\rm ph}$, and the adiabatic assumption
($\omega_{\rm ph} \ll E_{\rm F}$) breaks down.
In this context Migdal's theorem \cite{migdal}
does not apply and one needs to take
into account nonadiabatic effects not included in the Migdal-Eliashberg (ME)
theory of superconductivity.
Detailed studies have shown that the nonadiabatic contributions,
which are well represented by the vertex function, present
a complex momentum-frequency structure, in which small
${\bf q}$-scattering leads to an enhancement of the effective
superconducting pairing, while large-${\bf q}$ scattering leads to a reduction
of it.\cite{PSG,GPS,GPSprl}
The strong ${\bf q}$-modulation of the electron-phonon interaction
due to the electronic correlation is thus expected to give rise to
a net enhancement of the superconducting pairing.

The purpose of the present paper is twofold. On one hand we
wish to quantify the microscopic dependence of the screening
properties of a correlated system on relevant quantities
as the electron density of the Hubbard repulsion;
in addition we apply the derived screened electron-phonon interaction
to evaluate the role of the electronic correlation in the
context of the nonadiabatic superconductivity and to derive
a qualitative superconducting phase diagram.
To this aim
we introduce a model for the electronic Green's function of
the system, based on the decomposition of the total spectral function
in a coherent, itinerant part, and an incoherent localized
background corresponding to the Hubbard subbands. The relative
balance between the two parts varies as a function of doping and
electronic correlation. This will have important consequences on
the electronic screening and hence on the ${\bf q}$-modulation of the
effective electron-phonon scattering, as well as on the superconducting
properties.
We shall show that:

\begin{itemize}
\item the coherent excitations dominate the screening 
properties as well as the superconducting ones.

\item
the loss of coherent spectral weight approaching
half-filling is thus responsible for the reduction of the screening
properties and for the increase of the forward scattering
in the electron-phonon interaction.

\item in the strongly correlated regime
the selection of forward scattering  gives rise
to an enhancement of the effective electron-phonon interaction
within the context of the nonadiabatic superconductivity. These effects
however compete with the reduction of the quasi-particle spectral weight
which is detrimental for superconductivity.

\item the resulting phase diagram shares many similarities with
the one of the cuprates. In particular it shows
an overdoped region, where
superconductivity is suppressed
by negative nonadiabatic effects, an underdoped region, in which
superconductivity is destroyed by the loss of coherent
spectral weight, and an intermediate region in which the
predominance of small-${\bf q}$ scattering leads to an enhancement
of the nonadiabatic el-ph pairing which overcomes
the reduction of the coherent spectral weight.
\end{itemize}

We hereby wish to point out that a complete description of the
rich fenomenology of the cuprates is well beyond the aim
of the present paper. In particular, we shall not discuss,
for reason of simplicity, the symmetry of the gap, which
of course is of fundamental importance if one wishes to give a quantitative
description of these systems. We would like just to remark on this
point that a $d$-wave symmetry of the superconducting order parameter
was shown by many authors to naturally arise in the context of
a phonon pairing with a significant predominance of forward
scattering\cite{lichtenstein,abrikosov2,bouvier,chang}.
The competition between $s-$ and $d-$wave symmetry in a nonadiabatic
electron-phonon system was also studied in Ref. \onlinecite{paci}.
Taking into account explicitely the $d$-wave symmetry
of the gap would not change in a qualitative way the results
of the present work.

This paper is organized as follows: in section \ref{sec:model}
we introduce
our model Green's function; in section \ref{sec:ktf} we derive
an effective form for the electron-phonon interaction. In the
last section we write and solve the generalized Migdal-Eliashberg
equations, in the adiabatic and non-adiabatic limit, and discuss in detail
the competition of the different factors which determine the superconducting
critical temperature of our system.

\section{A model for correlated electron systems}
\label{sec:model}

As briefly discussed in the introduction, one of the main aims
of the present paper is to investigate how the screening properties
are affected by the presence of strong
 electronic correlation, and to parametrize
these effects in terms of microscopical quantities.
In particular we have in mind a Hubbard-like system where
itinerant electrons, with band dispersion $\epsilon_{\bf k}$
and bandwidth $E$,
interact each other through an onsite Coulomb repulsion $U$.
As we are going to see, a crucial role is played in this context
by the transfer of spectral weight as a function of the correlation
degree from low energy coherent states to the high energy
(Hubbard-like) incoherent ones.

In this section we present a simple, minimal model for the
electron spectral function which takes into account these
main effects and which can thus represent a proper starting point
to evaluate screening effects in correlated systems.

All the possible information about
the single-particle properties of the system is contained in  the
one-electron Green's function
$G({\bf k},\omega)$.
Without loss of generality we assume that the Green's function $G$
can be split in a coherent and an incoherent
contribution:\cite{abrikosovbook}
\begin{equation}
G({\bf k},\omega) =
G_{\rm coh}({\bf k},\omega)+G_{\rm inc}({\bf k},\omega),
\label{g1}
\end{equation}
where the coherent part $G_{\rm coh}$
describes the itinerant, quasi-particle like
properties of the electron
wavefunction,  while the incoherent part $G_{\rm inc}$
accounts for the incoherent
high energy excitations.
Due to its localized nature $G_{\rm inc}({\bf k},\omega)$
is only weakly dependent on the momentum quantum number,
so that the dependence on
${\bf k}$ can be reasonably neglected.

An important quantity which parametrizes the relative balance between
coherent and incoherent contributions is the
quasi-particle spectral weight $Z$, which is simply given by:
\begin{equation}
\int d\omega \frac{1}{\pi}\mbox{Im}
\left[ G_{\rm coh}({\bf k},\omega+i\delta)\right]
= Z,
\end{equation}
whereas the incoherent part obeys the sum rule:
\begin{equation}
\int d\omega \frac{1}{\pi}\mbox{Im}\left[
G_{\rm inc}({\bf k},\omega+i\delta) \right]= 1-Z.
\end{equation}
The quasi-particle spectral weight $Z$
can vary between 0 and 1, the two limits corresponding
to the insulating and metallic limit respectively.
It depends on the internal
parameters $U$ and $\delta$, where $\delta$ is the hole doping
($\delta=1-n$) and $n$ the total number of electrons ($n=1$ half-filled case).

Several techniques
have been developed to investigate the Hubbard model.\cite{emery}
Different starting points are employed according to whether main emphasis
has to be paid on the coherent (itinerant) or on the incoherent (localized)
features. For instance the so-called Hubbard I
approximation,\cite{hubbard} which is
exact in the atomic limit, is mainly aimed at a schematic representation
of the localized states, described by an upper
and a lower Hubbard band spaced by an energy gap of width $U$.
On the other hand the Gutzwiller technique\cite{gutzwiller}
and the mean field
slave bosons solution\cite{kotliar}
offer an useful tool to deal with the
coherent spectral weight of the electron Green's function: in this case
the quasi-particle spectral properties in the presence of strong correlation
are described in terms of an effective band
of non-interacting fermions
with spectral weight $Z$ and bandwidth $Z E$.

In this paper we introduce a new phenomenological model
to take into account in the simplest way and at the same level
the coherent and incoherent parts
of the Green's function.
We approximate the exact (unknown) coherent and incoherent parts
of $G({\bf k},\omega)$ in Eq. (\ref{g1}) respectively
with the Gutzwiller\cite{gutzwiller}
and Hubbard I \cite{hubbard}
solutions, namely:
\begin{equation}
G_{\rm coh}({\bf k},\omega)=
\frac{Z}{\omega -Z\epsilon_{\bf k} + \mu \pm i0^{+}},
\label{gcoh}
\end{equation}
\begin{eqnarray}
\label{ginc}
G_{\rm inc}(\omega)&=&
\frac{(1-Z)}{N_s}\sum_{\bf k}\left[\frac{(1-n/2)}{\omega-(1-n/2)
\epsilon_{\bf k} + \mu -U/2}\right. 
\nonumber
\\
&& +
\left.
\frac{n/2}{\omega-
(n/2)\epsilon_{\bf k} + \mu +U/2}\right],
\end{eqnarray}
where $\mu$ is the chemical potential,
$N_s$ is the total number of sites
and $Z$ is the quasi-particle weight
obtained in the Gutzwiller approximation in the paramagnetic
state at finite $U$ and
generic filling (Appendix \ref{appGutz}). Due to the localized nature
of the incoherent part we have replaced the $G_{\rm inc}({\bf k},\omega)$
given by the Hubbard I approximation with its momentum
average. Numerical calculations based on Dynamical Mean-Field Theory (DMFT)
confirm our qualitative picture of a spectral weight transfer from
a central coherent peak to a incoherent Hubbard-like background
with increasing $U$.\cite{georges}

\begin{figure}[t]
\centerline{\psfig{figure=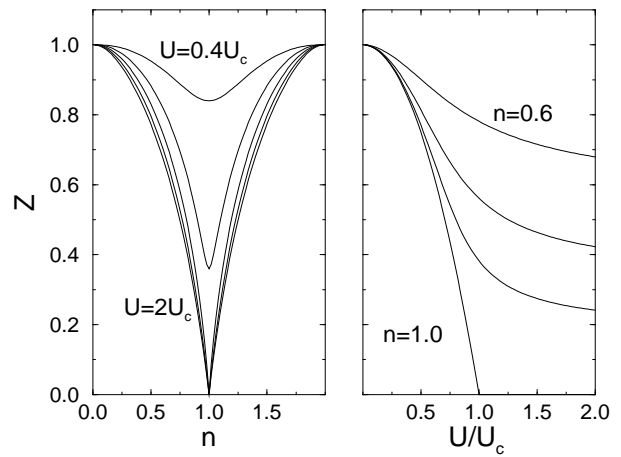,width=8cm,clip=}}
\caption{Quasi-particle spectral weight as determined by the Gutzwiller
solution at finite $U$ and $n$. Left panel: $Z$ as function of $n$ for
(from top to the bottom) $U/U_c = 0.4, 0.8, 1.2, 1.6, 2.0$. Right panel:
$Z$ as function of $U/U_c$ for (from top to the bottom)
$n = 0.6, 0.8, 0.9, 1.0$.}
\label{Zvsn-U}
\end{figure}

\begin{figure}[b]
\centering
\centerline{\psfig{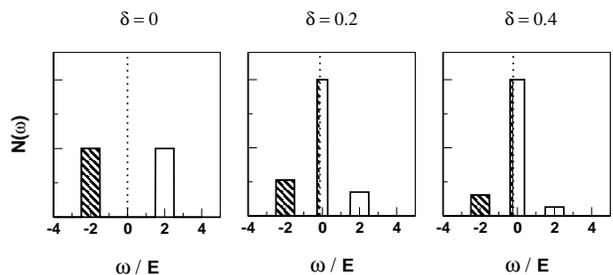}}
\caption{
\label{fig-dos}
Density of states
$N(\omega)=(1/\pi) \sum_{\bf k}
\mbox{Im} G({\bf k},\omega+i0^+)$ resulting from our
model, for $U=2U_c$ and different values of doping.
At half-filling the system is an insulator, and
its density of states is represented by two Hubbard-like features
at distance $U$ from each other; moving
away from half-filling a coherent peak starts forming,
with increasing weight $Z$. Dashed regions represent
filled states up to the chemical potential $\mu$ (dotted line).}
\end{figure}

The behaviour of $Z$ as function of the particles density $n$
and of the Hubbard energy $U$ is shown in Fig. \ref{Zvsn-U}.
The critical Hubbard energy $U_c$, which determines the
Brinkman-Rice transition at $n=1$ is related to the kinetic
energy $E_{\rm kin}$, which depends on the bare electron dispersion shape,
through the relation $U_c = 8 |E_{\rm kin}|$ .\cite{gutzwiller}
In the following we employ a bare constant density of states
(DOS) with
$N(\epsilon_{\bf k})=N_0=1/E$ for $\epsilon_{\bf k} \in [-E/2,E/2]$.
In this case, we have $U_c=2E$.
The chemical potential $\mu$ is determined by the total number of particles.
In Fig. \ref{fig-dos} we show
typical density of states $N(\omega)$
for the correlated system described by our model
[Eqs.~(\ref{gcoh})-(\ref{ginc})].

We would like to stress that the phenomenological model described by
Eqs.~(\ref{gcoh})-(\ref{ginc}) is not meant at all to be exhaustive of
the complex physics of a strongly correlated system.  In fact,
retardation effects are neglected, since we are assuming the
separation into two species of electrons to be independent of
frequency. More sofisticated methods of solution, including DMFT,
permit to treat the self-energy of a strongly correlated system in a
more careful way, retaining the correct frequency dependence of the
self-energy.

Our model has the advantage of being extremely simple and easy to
handle, and it allowed us to obtain explicit expressions for all the
relevant quantities of the coupled electron-phonon system; in
particular, we focus on the spectral weight transfer from the coherent
to the incoherent part of the Green's function when increasing the
degree of electronic correlation.  As we are going to see, this
feature will have important consequences on the electronic screening
and on the momentum dependence of the electron-phonon coupling.

\section{Screening and electron-phonon interaction}
\label{sec:ktf}

\subsection{Correlation effects on Thomas-Fermi screening}

The momentum dependence of the electron-phonon interaction usually plays
a marginal role in determining the electron-phonon
properties of common metals. The basilar reason for this is that
the bare long-range electron-phonon interaction is
effectively screened by the long-range Coulomb repulsion leading
to a weak momentum dependence.

The conventional Migdal-Eliashberg theory, which describes
electron-phonon effects both of the normal and superconducting states,
is formally derived starting from an effective electron-phonon
Hamiltonian, in which the Coulomb electron-electron repulsion does not
appear, apart from a weak residual electron-electron contribution in
the Cooper channel, $U_{{\bf k},{\bf q}}$, which gives rise the to the
Morel-Anderson ``pseudopotential'' term $\mu=N(0)U$.\cite{allen} The
physical quantities appearing in this effective Hamiltonian are thus
considered to have been already renormalized by the long-range Coulomb
interaction. In particular the electron-phonon matrix elements
$g_{{\bf k},{\bf k+q}}$ and the residual electron-electron repulsion
are usually considered to have a negligible momentum dependence, so
that the Eliashberg equations depend only on the frequency variables.

This drastic assumption works quite well in many conventional low
temperature superconductors with large carrier density since, in this
case, the long-range ${\bf q}$-dependence of the bare electron-phonon
and electron-electron interaction [$V({\bf q},\omega) \propto 1/|{\bf
q}|^2)$] is removed by the large metallic screening. This well-known
effect is usually expressed in terms of the (static) dielectric
function $\epsilon({\bf q})$, which in the RPA approximation reads:
\begin{equation}
\epsilon({\bf q}) = 1 + \frac{k_{\rm TF}^2} {|{\bf
q}|^2}, \label{scr}
\end{equation}
where $k_{\rm TF}$ is the Thomas-Fermi screening momentum defined
as
\begin{equation}
k_{\rm TF}^2 = - \lim_{{\bf q} \rightarrow 0} 4 \pi e^2
\Pi({\bf q},\omega=0),
\label{tfm}
\end{equation}
and $\Pi({\bf q},\omega)$:
\begin{equation}
\Pi({\bf q},\omega) = \frac{2}{N_s} \sum_{\bf k} \int d\omega' G({\bf
k+q},\omega+\omega') G({\bf k},\omega').
\label{bubble}
\end{equation}
The effective long-range interaction results thus
screened by conduction charge to give the Thomas-Fermi
expression:
\begin{eqnarray}
V_{\rm eff}({\bf q},\omega) &=&
\frac{V({\bf q},\omega)}{\epsilon({\bf q})}
\nonumber\\
&\propto& \frac{1}{|{\bf q}|^2 + k_{\rm TF}^2}.
\end{eqnarray}

In free electron systems the Thomas-Fermi vector is directly related
to the bare density of states via the simple relation
$\lim_{{\bf q} \rightarrow 0}
\Pi({\bf q},\omega=0) = - 2 N(0)$, where $N(0)$ is the density of states
per spin
at the Fermi level, so that $k_{\rm TF}^2= 8 \pi e^2 N(0)$.
In common metals, since $k_{\rm TF}$ is typically
larger than the Brillouin zone size ($k_{\rm BZ}$),
the effective (electron-electron, electron-phonon) interaction
$V_{\rm eff}({\bf q},\omega)$ can be
considered in first approximation
almost independent of the exchanged momentum ${\bf q}$.

Things are expected to be very different in correlated,
narrow band systems.
As we have mentioned before, strongly correlated
electrons, due to their reduced mobility,
are much less effective in screening
external perturbations, especially at small wavelengths.
For instance,
the reduction of the screening properties approaching a metal-insulator
transition in disorder alloys as well as in cuprates has been experimentally
signaled
in Refs. \onlinecite{osofskyprl,osofskyprb}.

In this section we employ
the simple model above introduced
for the description of the Green's function
to quantify the reduced screening properties
of correlated systems and their dependence on microscopic parameters,
such as the hole doping $\delta$ or the Hubbard repulsion $U$.
In order to do this, we compute the Thomas-Fermi vector
$k_{\rm TF}$, defined in Eq. (\ref{tfm}), using
the model described by
Eqs. (\ref{gcoh})-(\ref{ginc}) to evaluate the
RPA response function $\Pi({\bf q},\omega)$
according to Eq. (\ref{bubble}).
While higher order (vertex) diagrams are not taken into account
in this framework, we shall show that this simple model
is already sufficient to describe the reduction of
screening properties due to transfer of spectral weight from
the coherent to incoherent states.

Using Eqs. (\ref{gcoh})-(\ref{ginc})
the response function $\Pi$ can be written as a sum
of three different contributions:
\begin{equation}
\label{pb1}
\Pi = \Pi_{\rm c-c}+\Pi_{\rm c-i}+\Pi_{\rm i-i},
\end{equation}
where
the first one describes scattering processes which involve only coherent
states; the second term describes
scattering between the coherent peak and
the Hubbard lower/upper (incoherent) bands;
the last one describes processes which involve only localized
incoherent states in both the Green's functions of Eq. (\ref{bubble}).
In general  we expect that the total screening will be dominated
by the first contribution $\Pi_{\rm c-c}$ since the
itinerant coherent states are much more effective, because of their mobility,
in screening external perturbations
than the localized ones.

\begin{figure}[t]
\centering
\includegraphics[width=8cm]{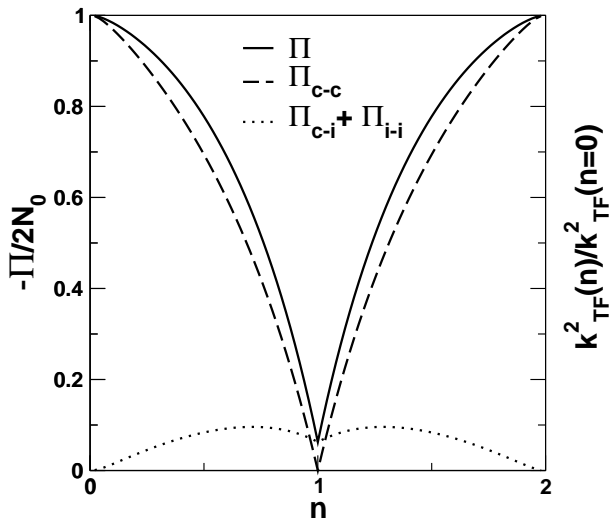}
\caption{\label{figRPA} Effective Thomas-Fermi screening $k^2_{\rm TF}$
(solid line) as a function of the electron density $n$ for
a correlated system described by our model
($U = 8 U_c$). The different contributions
to the total screening are also shown: coherent-coherent particle-hole
processes
(dashed line) and coherent-incoherent + incoherent-incoherent
contribution (dotted line).
Apart from half-filling, where the coherent contribution vanishes
and the screening is determined by the only residual incoherent
polarization, the static screening properties
of the system are dominated from the coherent quasi-particle excitations.
}
\end{figure}

In Fig. \ref{figRPA} we plot the RPA response function in units
of the bare DOS: $- \lim_{{\bf q}\rightarrow 0} \Pi({\bf q},\omega=0)/2N_0$,
as function of the electron filling.
Since for $n \rightarrow 0$ the screening properties are determined
only by the coherent part regardless any
correlation effects, this is also equivalent
to plot the Thomas-Fermi momentum $k_{\rm TF}^2$
as function of the electron density $n$: $k_{\rm TF}^2(n)/k_{\rm TF}^2(n=0)$.
As shown in figure
the net value of the Thomas-Fermi momentum is mainly determined
by the coherent-coherent excitations. Simple scaling considerations
show that the coherent-coherent contribution to the response
function is just equal to $\sqrt{Z}$ times
the Thomas-Fermi momentum of an uncorrelated system. The explicit expressions
of the other two terms are a bit more complicated
and they are reported in appendix \ref{appinc}.
Fig. \ref{figRPA} shows a drastic reduction of the screening properties
of the system as the metal-insulator
transition is approached at half-filling ($U>U_c$).
In this case the spectral weight of the coherent part is zero,
and the only residual small contribution to the
screening is due to incoherent excitations
which vanishes for $U \rightarrow \infty$.

\subsection{Poor screening and momentum dependence of
the electron-phonon interaction}

From Fig. \ref{figRPA} it is clear that the assumption of a
Thomas-Fermi momentum much larger than the exchanged phonon
momenta ${\bf q}$ breaks down as electronic correlation effects
get more and more relevant, namely approaching half-filling. In
this situation the effective electron-phonon interaction can be no
longer considered weakly dependent on ${\bf q}$ in the long-range
limit ${\bf q} \rightarrow 0$. On a microscopical ground the
screening of long-range Coulomb interaction renormalizes both the
bare electron-phonon matrix element $g_{\bf k,k+q}^0$ and the
phonon frequencies $\Omega_{\bf q}$. The
el-ph matrix element can be usefully written as
$g_{\bf k,k+q}^0 \simeq c(\Omega_{\bf q})/|{\bf q}|$, where
$c(\Omega_{\bf q})$ is a well behaved function of ${\bf q}$
in the limit $\lim_{{\bf q} \rightarrow 0}$ and it mainly depends
on the phonon frequency $\Omega_{\bf q}$.
If both the screening effects on $g_{\bf k,k+q}^0$ and
$\Omega_{\bf q}$ are
properly taken into account\cite{mahan} one can get an expression for the
effective total electron-phonon interaction:
\begin{equation}
V^{\rm el-ph}_{\rm eff}({\bf q},\omega) = \frac{c^2(\omega_{\bf q})}
{|{\bf q}|^2\epsilon({\bf q})}
D_{\bf q}(\omega),
\label{veff}
\end{equation}
where both the phonon propagator $D_{\bf q}(\omega)$
and the coupling function $c(\omega_{\bf q})$ are written
in terms of the screened phonon frequency $\omega_{\bf q}$.
Eq. (\ref{veff}) shows that
the long-range behaviour of the total el-ph interaction
$\propto 1/|{\bf q}|^2$,
when written as function of the screened phonon frequency,
is correct by the dielectric function $\epsilon({\bf q})$.

For an optical mode, $\omega_{\bf q}$ is only weakly dependent on
${\bf q}$ and the leading dependence on ${\bf q}$ of
Eq. (\ref{veff}) comes from the term
$\propto  1/[\epsilon({\bf q})|{\bf q}|^2]$.
These screening effects can be conveniently dealt with
by introducing the screened el-ph matrix element
$g_{\bf q}$:
\begin{equation}
g_{\bf q}^2 = \frac{|g_{\bf q}^0|^2}{\epsilon({\bf q})}
\propto 
\frac{1}{|{\bf q}|^2+k_{\rm TF}^2}.
\label{qeff}
\end{equation}
The el-ph scattering is thus roughly described (we remind these expression
were derived in the limit ${\bf q} \rightarrow 0$)
by a lorentzian function in the space $|{\bf q}|$.
It is also useful to introduce the dimensionless variables
$Q=|{\bf q}|/2k_{\rm F}$ and $Q_c=k_{\rm TF}/2k_{\rm F}$, so that:

\begin{equation}
\label{gdrQ}
|g(Q)|^2 \simeq g^2
\frac{1}{Q^2+Q^2_c}.
\end{equation}
The parameter $Q_c$ represents a cut-off for the exchanged
phonon momenta: the electron-phonon scattering will be
operative for $Q \lsim Q_c$, and negligible
for $Q \bsim Q_c$.

The momentum structure resulting in Eq. (\ref{gdrQ}) plays a crucial role
in the Cooper pairing in the coherent-coherent channel where
the momentum is a good quantum number.
For these contributions
the total strength of the electron-phonon coupling is linked with
the momentum average of Eq. (\ref{veff}) over the Fermi surface.
For a isotropic system,
using polar coordinates $\int d\Omega = \int_{0}^{2\pi} d\phi
\int_0^1 d\cos \theta$ and reminding that
$Q = \sin (\theta/2)$,
we obtain:
\begin{eqnarray}
\label{gveraFS}
\left<|g(Q)|^2\right>_{\rm FS}
&=& \frac{\int d\phi   \int_0^1 QdQ\frac{\displaystyle g^2}
{\displaystyle Q^2+Q^2_c}}{\int d\phi   \int_0^1 Q dQ}
\nonumber\\
&=& g^2 \ln\left(\frac{1+Q^2_c}{Q^2_c}\right).
\end{eqnarray}
In common metals $Q_c \sim 0.5-1$ so that
$\ln\left[(1+Q^2_c)/Q^2_c\right]$ is of the order of $1$.
On the other hand, in poorly screened systems $Q_c \ll 1$ and
the resulting el-ph coupling is sensibly enhanced. In the following
we shall consider $Q_c \simeq 0.7$ as representative case
of uncorrelated usual metals.

For practical purposes,
following Refs. \onlinecite{GPS,GPSprl},
we approximate the lorentzian behaviour of Eq. (\ref{gdrQ})
with a
Heaviside $\theta$ function:
\begin{equation}
\label{gtheta}
|g(Q)|^2 \rightarrow g^2 \alpha\, \theta(Q_c-Q).
\end{equation}
In order to preserve in this mapping the total strength
of the el-ph coupling, the prefactor $\alpha$ has to be
determined by requiring
the resulting el-ph coupling strength, namely
the average of $g^2$
over the Fermi surface, to be equal
for Eqs. (\ref{gdrQ}) and (\ref{gtheta}).
With this condition we find:
\begin{equation}
\label{gvera}
|g(Q)|^2=g^2 \frac{1}{Q_c^2}\ln\left(\frac{1+Q^2_c}{Q^2_c}\right)
\theta(Q_c-Q),
\end{equation}

As a final remark of this section we note that
the momentum dependence of $|g(Q)|^2$ is not expected
on the other hand to be effective in the incoherent-coherent and
incoherent-incoherent contributions to the electron-phonon interaction,
where the exchanged momentum ${\bf q}$ is no more a good quantum
number.
In this case the effective incoherent electron-phonon coupling
is roughly given by its momentum average on the Brillouin zone,
which we shall set in the following to be equal to $g^2$.

\section{Generalized Migdal-Eliashberg equations}

In the previous sections we have introduced a simple model for
an electron-phonon system in the presence of electronic correlation.
In particular we have reduced, in an approximate way, the complex
problem of the interplay between electron-phonon and electron-electron
interactions to a purely electron-phonon
system described by
an effective one-particle Green's function
[Eqs. (\ref{g1}), (\ref{gcoh}), (\ref{ginc})]
and an effective electron-phonon matrix element
$g(Q)$ [Eq. (\ref{gvera})]. After this mapping, the Baym-Kadanoff
theory\cite{baym}
assures that the functional form of the superconducting
equations will be the same of a purely electron-phonon system:
\begin{eqnarray}
\label{BaymTc}
\Phi &=& \Phi_{\rm el-ph}[g,G,\Phi],
\\
Z&=&Z_{\rm el-ph}[g,G,\Phi],
\label{BaymZ}
\end{eqnarray}
where $\Phi$ is the superconducting order parameter;
the Green's function $G$ and the matrix element $g$ are defined
by Eqs. (\ref{g1}), (\ref{gcoh}), (\ref{ginc}), (\ref{gvera}),
as mentioned above.
In order to obtain an explicit expression for
Eqs. (\ref{BaymTc})-(\ref{BaymZ})
we should specify {\em in which framework} we are going to treat the electron
phonon interaction.
In particular, we observe that the conventional ME
theory, in particular, is based on the assumption that the phonon frequencies
are much smaller than the electronic Fermi energy, $\omega_{\rm ph}
\ll E_{\rm F}$ ( adiabatic limit).
This theory works quite well in the conventional
low temperature superconductors, where no electronic correlation
is present and $E_{\rm F}$ is of the order of $5-10$ eV.
On the other hand, the strong band renormalization in
correlated systems described in Sec. \ref{sec:model} questions
the adiabatic assumption, especially as, approaching half-filling,
 the renormalized bandwidth $\sim Z E$ can be comparable with
$\omega_{\rm ph}$.
In these systems a more suitable description can be obtained
in the framework of the non adiabatic theory of
superconductivity.\cite{PSG,GPS,GPSprl}
Eqs. (\ref{BaymTc})-(\ref{BaymZ}) can
be rewritten as:
\begin{eqnarray}
Z_n& = &1 + \frac{T_{\rm c}}{\omega_n}\sum_m
\Gamma_Z ([G];\omega_n,\omega_m)
\eta_m[G] ,
\label{z}\\
\Phi_n& =&  T_c \sum_m
\Gamma_\Phi([G];\omega_n,\omega_m)
\frac{\Phi_m}{\omega_m Z_m}\eta_m^\Delta[G],
\label{gap}
\end{eqnarray}
where the electron-phonon kernels $\Gamma_Z ([G];\omega_n,\omega_m)$ and
$\Gamma_\Phi([G];\omega_n,\omega_m)$ contain the nonadiabatic
vertex ($P$) and cross ($C$)
contributions to the self-energy and to the Cooper pairing channels:
\begin{eqnarray}
\Gamma_Z ([G];\omega_n,\omega_m)& = &\lambda_{n-m}
\left[ 1 + \lambda P([G];\omega_n,\omega_m,Q_c) \right],
\nonumber\\
\Gamma_\Phi ([G];\omega_n,\omega_m)& = &\lambda_{n-m}
\left[1 + 2\lambda P([G];\omega_n,\omega_m,Q_c)\right]
\nonumber\\
&& + \lambda^2 C([G];\omega_n,\omega_m,Q_c) - \mu.
\nonumber
\end{eqnarray}
Here $\lambda_{n-m}$ is linked with the electron-phonon spectral function
$\alpha^2\!F(\omega)$
through the relation
$\lambda_{n-m}=2\int\! d\omega\, \alpha^2\!F(\omega)\omega/
[\omega^2+(\omega_n-\omega_m)^2]$,
$\lambda=\lambda_{n-m=0}$ and $\mu$ is the short-range residual
electron-electron repulsion.
The breakdown of the adiabatic hypothesis determines the need for
the explicit inclusion of the vertex ($P$) and cross ($C$) functions in
Eqs. (\ref{z})-(\ref{gap}) and it affects the expression
of $\eta_m[G]=\sum_{\bf k}G({\bf k},\omega)$
and $\eta_m^\Delta[G]=\sum_{\bf k}G({\bf k},\omega)G({\bf -k},-\omega)$
through finite bandwidth effects.
The momentum dependence of the superconducting equations
has been averaged on the Fermi surface and it gives rise
to the strong dependence on $Q_c$ in the vertex and cross terms.
In Eq. (\ref{z})-(\ref{gap}) we have moreover implicitly expressed the
functional dependence of the electron-phonon kernels $\Gamma_Z$,
$\Gamma_{\Delta}$ as well as of the quantities $P$, $C$ and $\eta$,
on the Green's function $G$ which we remind is modeled as in
Eqs. (\ref{g1}), (\ref{gcoh}), (\ref{ginc}).

\begin{figure}[t]
\centerline{\psfig{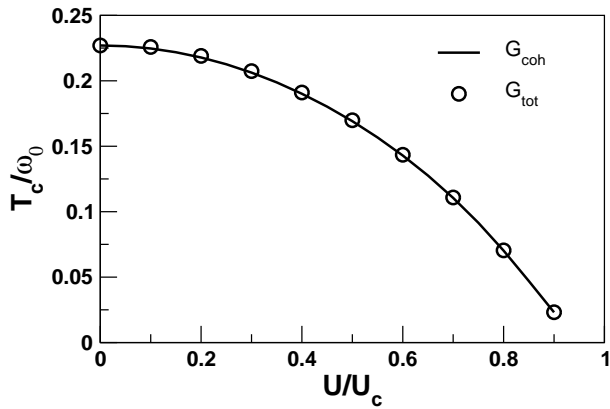}}
\caption{Comparison of the critical temperature $T_c$ as a
function of $U$ for the half-filling case,
using the full integral kernel in Eqs. (\ref{z})-(\ref{gap}) (empty circles) 
and using only its coherent part (solid line).}
\label{tcu-fig}
\end{figure}

Before solving Eqs.~(\ref{z})-(\ref{gap}) in the whole range of
doping, we would like to discuss the different role of the coherent
(itinerant) states and the incoherent (localized) states, described
respectively by Eqs. (\ref{gcoh}), (\ref{ginc}) on the superconducting
properties.  As we have seen in Sec. \ref{sec:ktf}, the electronic
screening is mainly dominated by the
coherent term of the electronic Green's function, which describes mobile
electrons for which {\bf k} is a good quantum number.

Similar considerations can be made also for superconductivity: we
expect, in fact, that the coherent electrons, which have a high
mobility, will give a more relevant contribution to the
superconducting critical temperature.  To check the validity of this
hypothesis, we have solved Eqs.~(\ref{z})-(\ref{gap}) in the ME limit
({\em i.e.}, neglecting vertex corrections), once using an integral
kernel containing the full Green's function (Eq. \ref{g1}), and once
an integral kernel with only the coherent part of the Green's function
[Eq. (\ref{gcoh})], as a function of the Hubbard repulsion $U$.  In
Fig.~(\ref{tcu-fig}), we show as empty circles the results obtained
with the full kernel, and with solid line the critical temperature
obtained using only the coherent part.  The two sets of data are
hardly distinguishable, pointing out that the increase of $T_c$ due to
the coherent-incoherent and incoherent-incoherent couplings is
negligible.

After this observation, in the following
the functional dependence on the total Green's
function $G$ in Eqs. (\ref{z})-(\ref{gap}) can be in good
approximation replaced by the only coherent part, explicitly:
$\Gamma_Z[G_{\rm coh}]$, $\Gamma_\Phi[G_{\rm coh}]$, $P[G_{\rm coh}]$,
$C[G_{\rm coh}]$, $\eta_m[G_{\rm coh}]$.  As we show in Appendix
\ref{appinc}, when the reduced spectral weight and bandwidth are taken
into account, this corresponds to a proper rescaling of the analytical
expressions for these quantities evaluated in the absence of
correlation in Refs. \onlinecite{sgp,cgp}.

It is interesting to compare our model
with the two-band superconductivity \cite{suhl},
which has recently driven a considerable attention due to MgB$_2$. 
\cite{kortus}. In that case, the opening of inter-band scattering
channels leads to an enhancement of the critical temperature.
For some respects, our model could also be seen as an effective
two-band system, made up of a very narrow band of mobile electrons 
and another band of localized electrons, coupled to each other. 
However, we note that, since the spectral weight of each single
band is not conserved, the onset of the high-energy bands of localized
electrons is accompanied by a decrease of the quasi-particle spectral
weight, resulting in an effective reduction of the Cooper pairing.
\subsection{Doping effects and phase diagram
of the nonadiabatic superconductivity}

Eqs. (\ref{z})-(\ref{gap}) represent our  tool to investigate
the loss of the superconducting properties due to the electronic
correlation approaching half-filling. We can in fact evaluate all
the relevant quantities, such as the
electron-phonon interaction kernels $\Gamma_Z$, $\Gamma_\Delta$,
the electron Green's function $G$,
the vertex and cross functions $P$, $C$, and the momentum cut-off
$Q_c$ as a function of the microscopic parameters as the
hole doping $\delta$ and the Hubbard repulsion $U$.
We shall show that the phase diagram as a function of the doping
is governed by two competing effects: one driven
by the reduction of the coherent spectral weight approaching half-filling,
which is detrimental for superconductivity, and the other by the complex
behaviour of the non-adiabatic terms, which increase the effective
pairing as $\delta \to 0$ and decrease it as $\delta \to 1$.

Since we are mainly interested in the region $\delta \to 0$ of
the phase diagram, we disregard for simplicity the analytical
dependence of the non-adiabatic terms on the chemical
potential.
The behaviour of the ``bare'' $P$ and $C$ as a function of doping is in fact
determined by the density of electrons ($n=1-\delta$);
this dependence is much weaker than the dependence of $Z$ and $Q_c$
close to half-filling (see Figs. \ref{figRPA}, \ref{Zvsn-U}).

\begin{figure}[t]
\centerline{\psfig{figure=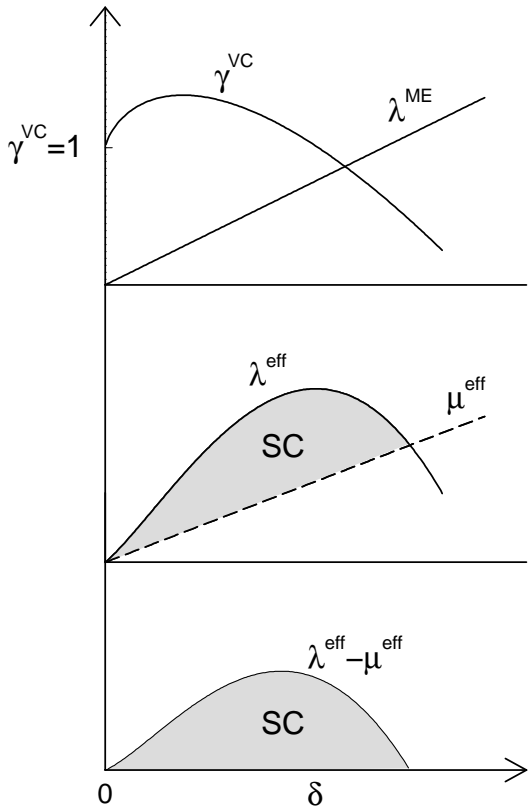,width=7cm,clip=}}
\caption{Graphical sketch of the different contributions to the
effective superconducting coupling. Top panel:
the coupling function $\lambda^{\rm ME}$
is mainly determined by the coherent spectral weight,
and it exhibits a monotonous
growing behaviour as a function of doping.
The vertex factor $\gamma^{\rm VC}$
tends to enhance the effective coupling at low doping and to depress it
at high doping. Middle panel: the total effective electron-phonon
coupling $\lambda^{\rm eff}= \lambda^{\rm ME} \gamma^{\rm VC}$
has a maximum at some
finite value of $\delta$; when the effective Morel-Anderson pseudopotential
is subtracted, superconductivity is suppressed at high doping.
Lower panel: resulting phase diagram for superconductivity:
superconductivity is only possible in a finite region
of phase space (gray region), where $\lambda^{\rm eff}-\mu^{\rm eff}$ is
positive.}
\label{figcoup}
\end{figure}

Before solving Eqs. (\ref{z})-(\ref{gap}) numerically
to obtain the critical temperature $T_c$ as a function of doping,
we wish to discuss the phase diagram of our model
in terms of
simple intuitive physical arguments, based on an effective electron-phonon
coupling.
Let us consider for the moment the electron-phonon interaction alone,
without any residual Coulomb
repulsion, namely $\mu=0$.
Eq. (\ref{gap}) can be rewritten in a simplified way as:
\begin{equation}
\Phi_n  \simeq  T_{\rm c}\sum_m
Z\lambda \left[ 1 + 2 Z \lambda P + Z \lambda C \right]
K_{n-m} \Phi_m,
\label{gap2}
\end{equation}
where
we have simplified, according Appendix \ref{appinc},
the main dependences on $Z =Z(U,\delta)$
of each quantity.
In this way, we can
roughly see the total electron-phonon
coupling as the
product of two terms: an effective electron-phonon
coupling of ME theory renormalized by the electronic
correlation, $\lambda^{\rm ME}$, and the enhancement
due to nonadiabatic vertex and cross (VC) diagrams
$\gamma^{\rm VC}$:
\begin{eqnarray*}
\lambda^{\rm eff}&=&\lambda^{\rm ME} \gamma^{\rm VC},
\\
\lambda^{\rm ME}&=&Z \lambda,
\\
\gamma^{\rm VC} &=& 1 + 2 Z \lambda P(Q_c) + Z \lambda C(Q_c).
\end{eqnarray*}
The schematic behaviour of these quantities as a function of the
hole doping $\delta$ is shown in the
upper panel of Fig. \ref{figcoup}.
The physics behind the $\delta$-dependence of $\lambda^{\rm ME}$
can be easily related
to the loss of spectral weight
approaching the metal-insulator transition for $\delta \rightarrow 0$.
This effect, which is present also in $\gamma^{\rm VC}$, is however
in that case competing with the enhancement of the effective coupling due
to $P(Q_c)$
and $C(Q_c)$ which will be maximum and positive
close to half-filling
(where $Q_c \rightarrow 0$) and negative at high dopings.
The interplay between these two effects will give rise to a maximum
of $\gamma^{\rm VC}$, and hence of $\lambda^{\rm eff}$, somewhere
in the small doping region where the competition between the spectral
weight loss and the positive nonadiabatic effects is stronger
(see upper and middle panels in Fig. \ref{figcoup}).

We can now also consider the effect of the residual Morel-Anderson-like
repulsion $\mu$; first of all, we observe that the reduction of
spectral weight will lead to an
effective repulsion $\mu^{\rm eff} \simeq Z \mu$.
Superconductivity will be possible only when the net electron-phonon
attraction overcomes the repulsion term:
$\lambda^{\rm eff}-\mu^{\rm eff} > 0$ (see lower panel of Fig.
\ref{figcoup}).
The resulting total coupling is expected to exhibit a
``bell'' shape which is mostly due to the $\delta$-dependence
of the nonadiabatic factor $\gamma^{\rm VC}$. It is interesting
to note two things. First, in the extreme case
$\lambda^{\rm ME} \lsim \mu^{\rm eff}$, where no superconductivity would be
predicted in the whole $\delta$ range by the conventional ME
theory, we could expect finite $T_c$ in a small $\delta$ region, due
to purely nonadiabatic effects $\lambda^{\rm eff}=
\lambda^{\rm ME} \gamma^{\rm VC} > \mu^{\rm eff}$. Secondly, it is clear
that within the ME framework a net attractive interaction in
the Cooper channel at a certain
doping $\delta$, which corresponds to $\lambda^{\rm ME} > \mu^{\rm eff}$,
would imply a superconducting order also at larger $\delta$ since
the two quantities
$\lambda^{\rm ME}, \mu^{\rm eff}$ scale in the same way $\propto Z$;
on the other hand, in the nonadiabatic
theory superconductivity, $T_c$ is expected to be limited
to some maximum value of doping, due to the negative contribution
of the nonadiabatic diagrams $P$ and $C$ at large $\delta$ (large $Q_c$'s).

We can now quantify the simple arguments discussed so far.
A quantitative estimate of the strength of the superconducting pairing
is given by the highest eigenvalue $v^{\rm max}$
of the superconducting integral
kernel
in Eq. (\ref{gap}), computed at low $T$;
at a given temperature $T$  and doping $\delta$
superconductivity
occurs if $v^{\rm max} \ge 1$ and the superconducting pairing
(and $T_c$) is stronger as $v^{\rm max}$ is larger.

\begin{figure}[t]
\centerline{\psfig{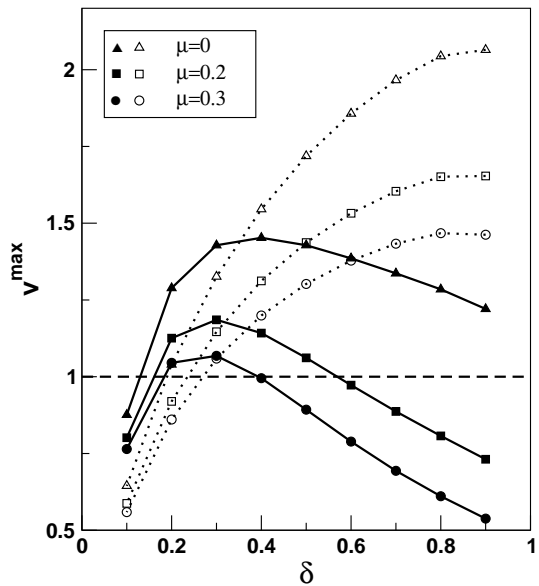}}
\caption{Maximum eigenvalue $v^{\rm max}$
as a function of doping, evaluated at $T=0.01\omega_0$
for different values of $\mu$, with $\lambda =1$,
$\omega_0 = 0.8 E/2$, and $U = 8 U_c$.
Empty symbols (dashed lines) represent ME theory, filled
symbols (solid lines)
the nonadiabatic theory described by Eqs. (\ref{z})-(\ref{gap}).
}
\label{figvmax}
\end{figure}

\begin{figure}[t]
\centerline{\psfig{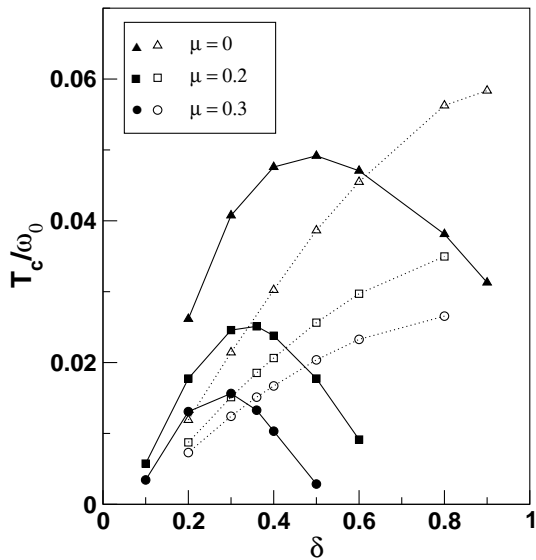}}
\caption{Phase diagram of $T_c$ vs. $\delta$ for
the adiabatic (empty symbols, dashed lines) and
nonadiabatic (filled symbols, solid lines) theory. Details
as in Fig. \ref{figvmax}}
\label{tcvert2}
\end{figure}

In Fig. \ref{figvmax} we compare the behaviour of  $v^{\rm max}$
as a function of $\delta$, obtained at $T=0.01\omega_0$ using an
Einstein spectrum
for different values of $\mu$ in ME (open symbols, dashed lines)
and in the nonadiabatic theory (full symbols, solid lines).
The Hubbard repulsion was set at $U = 8 U_c$ and the
phonon frequency at $\omega_0 = 0.8 E/2$, where
$E/2$ is the bare half-bandwidth (unrenormalized by correlation effects).
The corresponding phase diagram $T_c$ vs. $\delta$ is reported in
Fig. \ref{tcvert2}.
In agreement with our previous discussion in the ME framework
$v^{\rm max}$ and $T_c$ decrease monotonously as the hole
doping is reduced. On the other hand, the
corresponding results in the
nonadiabatic theory display a more complex behaviour, showing
that the effective nonadiabatic pairing is larger
than the ME one at low doping and smaller at high doping.

As we have already discussed, the bell-shape of the
highest eigenvalue $v^{\rm max}$ and of the critical temperature $T_c$
can be attributed to the dependence of the magnitude and sign of the
nonadiabatic terms on $Q_c$, which, in turn, strongly depends on
doping. For high doping the nonadiabatic
contributions are negative and decrease
$v^{\rm max}$ and $T_c$ with respect to their ME values.
Decreasing $\delta$ the nonadiabatic terms turn from negative
into positive and $v^{\rm max}$ and $T_c$ increase up to a maximum value.
As the hole doping is further decreased ($\delta \to 0$), the
loss of spectral weight becomes the dominant effect and it finally
leads to the complete suppression of superconductivity.
The inclusion of the residual Coulomb repulsion  in the
Cooper channel, $\mu$, leads to an overall reduction of
the superconducting
pairing. The effect is more pronounced in the nonadiabatic
theory than in ME, since in this case a very small
value of $\mu$ is enough to suppress superconductivity
in a large region of phase space at high dopings.

\section{Discussion and Conclusions}

The main aim of the present paper is the description, on microscopical grounds,
of the hole doping dependence of the electron-phonon and superconducting
properties of a strongly correlated system
within the context of a nonadiabatic electron-phonon theory.
The need for a nonadiabatic treatment of the electron-phonon interaction
in correlated systems comes from the fact that, as a metal-insulator
transition is approached, the electronic bandwidth is strongly
reduced, and the adiabatic assumption $\omega_{ph}/E_{\rm F}$ on
which Migdal's theorem is based breaks down.

Past studies have shown that the inclusion of nonadiabatic effects
can lead to a strong enhancement or depression of $T_{\rm c}$ depending
on the value of the exchanged momenta and frequencies:
if a microscopic mechanism leads to a predominance of the
forward scattering in the electron-phonon interaction, $T_{\rm c}$
is strongly enhanced.
This effect was schematized in the past with the introduction
of an effective cut-off in the electron-phonon interaction ($Q_{\rm c}$),
which was argued to be due to strong correlations effects.

In this work we have related the existence of $Q_{\rm c}$ with the reduction
of the screening properties due to correlation
 of a metal approaching a metal-insulator transition.
The same effects which are responsible for the reduction of the
screening (namely the loss of {\bf k}-space coherence) are however also
strongly detrimental to superconductivity. In this work we
have analyzed how the interplay between these effects
is reflected on a $T_{\rm c}$ vs doping phase diagram.

We have introduced a simple analytical model to simulate the effects
of the strong correlation on the one electron Green's function.
This model has also been employed to estimate the role
of the electronic screening on the electron-phonon
scattering in correlated systems. We have shown that
the reduction of the metallic character due to the electronic correlation
implies a reduction of the ``effective'' Thomas-Fermi screening
approaching $\delta = 0$, where correlation is stronger.
This results in a predominance
of forward (small-${\bf q}$) scattering, which has been parametrized
in terms of a phonon momentum cut-off $Q_c= k_{\rm TF}/2k_{\rm F}$,
where $k_{\rm TF}$ and $k_{\rm F}$
are respectively the Thomas-Fermi and the Fermi vectors.

We have also shown how the different parts of the electronic Green's
function contribute to the superconducting pairing; in particular,
we have shown that the superconducting critical temperature
is mainly determined by the coherent excitations.
The similarities and differences between our model and the two-band
superconductivity \cite{suhl,kortus} have also been discussed.

Solving the nonadiabatic generalized ME equations,
we obtained  a
$T_c$ vs. $\delta$ diagram, which
can be ideally divided into three regions:
\begin{enumerate}
\item[($a$)] a high doping region,
where superconductivity is suppressed by
the negative contribution of the nonadiabatic channels to the electron-phonon
pairing;
\item[($b$)] an extremely low doping region, where the poor metallic character
is reflected in a vanishing coherent spectral weight.
In this region,
superconductivity
is extremely unstable and it can be overwhelmed by
other electronic or structural instabilities induced by
spin and charge degrees of freedom (antiferromagnetic fluctuations,
stripes, charge-density-waves, pseudogaps, \ldots).
\item[($c$)]
an intermediate doping region, in which the loss of coherent spectral
weight
is not large enough to prevent superconductivity,
which is in turn enhanced by
the positive contribution of the nonadiabatic channels of interactions.
\end{enumerate}

The resulting phase diagram bares strong resemblance with that
of cuprates.
We have
in fact an overdoped region,
where superconductivity is triggered on by the positive contribution of
the nonadiabatic channels as doping is decreased; an optimal
doping, where the enhancement due to the nonadiabatic interaction
is counterbalanced by the reduction of the metallic character, and an
underdoped region, where superconductivity disappears due to
the incipient metal-insulator transition.
In the qualitative scenario outlined here
the origin of superconductivity in cuprates can be understood
by focusing on the overdoped region, where the materials retain
defined metallic properties; on the other hand, the exotic phenomenology of the
underdoped region is only marginal. The occurrence of different kinds
of electronic/structural instabilities,
not discussed in the present paper,
is thus thought to be a by-product of
the loss of metallic character which also drives
the {\em suppression}
of $T_c$ as $\delta \rightarrow 0$
more than to be the secret of the superconducting pairing.

Once more, we wish to stress that what we present in this paper is a
general scenario, based on the microscopical description of the
interplay between nonadiabatic effects and strong electronic
correlation. A quantitative understanding of the specific phase
diagram of cuprates should of course take into account specific
features of these materials, such as Van Hove singularities and the
$d$-wave symmetry of the gap.  The possibility of a $d$-wave pairing
within the context of electron-phonon superconductivity has been
discussed elsewhere \cite{paci}; we remind here that $d$-wave symmetry
was shown to be favoured by forward scattering, which in our model is
enhanced as $\delta \to 0$.

The qualitative behaviour of our results would be however left
unchanged by the inclusion of these effects.

On the experimental ground we observe that the present scenario is
supported by a detailed
analysis of $T_c$ vs. normal state properties in different families
of cuprates. In Ref. \onlinecite{osofskyprb}, for example,
 the complex  behaviour of $T_c$
approaching the metal-insulator transition either
by reducing the doping or by increasing the disorder
was nicely pointed out by Osofsky {\em et al.}.
The relation between $T_c$ and reduced screening properties was
also discussed there. Although we do not attempt to discuss
the scaling relations close to the metal-insulator transition in region
($b$), where a more specific treatment of the electronic correlation is needed,
we think our analysis is somehow complementary to that of
Ref. \onlinecite{osofskyprb}.
This scenario can also open new perspectives on the
remarkable increase of $T_c$ in granular metals and alloys.\cite{osofskyprl}

Furthermore,
a strong doping dependence of the electron-phonon properties in cuprates
has also been reported by inelastic
X-ray measurements of the phonon dispersion.\cite{dastuto}
Experimental data in NCCO
compounds were shown to be compatible with the
theoretical calculations, based on the shell model,
 assuming a negligible Thomas-Fermi vector
($Q_c=0$) for the strongest correlated undoped compound
($\delta=0$),
whereas a Thomas-Fermi momentum $k_{\rm TF} \simeq 0.39 \AA^{-1}$,
comparable to that for LSCO, was estimated for $\delta \simeq 0.14$.
The corresponding dimensionless cut-off $Q_c$ would be hence
estimated
$Q_c \simeq 0.26$ by using an in-plane
Fermi vector $k_{\rm F}^{ab} \simeq 0.74 \AA^{-1}$.

L.B. wishes to thank Massimo Capone and Giorgio Sangiovanni for
useful discussion and comments.
The authors acknowledge financial support
by the MIUR projects COFIN2001 and FIRB-RBAU017S8R, and by
the INFM project PRA-UMBRA.

\begin{appendix}
\section{Gutzwiller solution for generic $U$ and $n$}
\label{appGutz}

In this appendix we provide a brief overview about the analytical
solution of the Gutzwiller approximation for generic filling
and Hubbard repulsion.

Let us write
the Hubbard Hamiltonian within the Gutzwiller
approximation (in the paramagnetic state) as:
\begin{equation}
H = - \gamma(U,n,d) |\bar{\epsilon}| + U d,
\label{ham-gutz}
\end{equation}
where $n$ is the total electron filling, $d$ the density of
double occupancy sites, $\bar{\epsilon}$ the kinetic energy
for site, and
\begin{equation}
\gamma(U,n,d) = \frac{2(n-2d)}{n(2-n)}
\left[\sqrt{1-n+d}+\sqrt{d}\right]^2.
\end{equation}

Minimizing Eq. (\ref{ham-gutz}) with respect to $d$ yields:
\begin{eqnarray}
0 &= &\left[\sqrt{1-n+d}+\sqrt{d}\right]^2
-\left(\frac{n}{2}-d\right)
\frac{\left[\sqrt{1-n+d}+\sqrt{d}\right]^2}
{\sqrt{d(1-n+d)}}
\nonumber\\
&&+ 2\left(\frac{U}{U_c}\right)n(2-n) ,
\end{eqnarray}
where we have introduced as usual $U_c = 8 |\bar{\epsilon}|$.

After expanding the squares $\left[\sqrt{\ldots}+\sqrt{\ldots}\right]^2$
one can now isolate on the right side the remaining square roots:
\begin{eqnarray}
&&2 n (2-n)\left(\frac{U}{U_c}\right) + (1-2n+4d)
\nonumber\\
&=&\left(\frac{n}{2}-d\right)
\frac{(1-n+2d)}{\sqrt{d(1-n+d)}}
- 2 \sqrt{d(1-n+d)},
\label{pr1}
\end{eqnarray}
and, by squaring both the sides of Eq. (\ref{pr1}),
all the remaining square roots are removed and we are left
with a third order polynomial expression for $d$. We obtain namely:
\begin{equation}
A_3 d^3 + A_2 d^2 + A_1 d + A_0 = 0,
\label{eqgutz}
\end{equation}
where
\begin{eqnarray}
A_3 &=& 16 n(2-n) \left(\frac{U}{U_c}\right),
\\
A_2 &=& 4 n(2-n) \left(\frac{U}{U_c}\right)
\left[ n(2-n) \left(\frac{U}{U_c}\right) - 6n +5 \right],
\\
A_1 &=& (1-n)
\Bigg[
4n^2(2-n)^2\left(\frac{U}{U_c}\right)^2
\nonumber\\
&& \hspace{1.5cm}
+ 4 n(2-n)(1-2n)\left(\frac{U}{U_c}\right) -n\Bigg],
\\
A_0 &=& - \frac{n^2(1-n)^2}{4}.
\end{eqnarray}
Eq. (\ref{eqgutz}) can be easily solved to obtain $d_{\rm min}$,
and, in the standard notations, the Gutzwiller factor
$Z(U,n) = \gamma(U,n,d_{\rm min})$.

\section{Analytical expression of different physical quantities}
\label{appinc}

\subsection{Thomas-Fermi Screening}
In this section we provide some useful analytical expressions
for the different contributions ($\Pi_{\rm c-c}$,
$\Pi_{\rm c-i}$, $\Pi_{\rm i-i}$)
to the response function $\Pi$
involved in the evaluation of the Thomas-Fermi screening
as limit $k_{\rm TF}^2 \propto 
\lim_{{\bf q} \to 0}\Pi({\bf q},\omega=0)$.

In the RPA approximation
$\Pi({\bf q},\omega)$ is given by:
\begin{equation}
\Pi({\bf q},\omega) = \frac{2}{N_s} \sum_{\bf k} \int d\omega' G({\bf
k+q},\omega+\omega') G({\bf k},\omega').
\label{bubble2}
\end{equation}
We employ the simple model of Eqs. (\ref{g1}),
(\ref{gcoh}), (\ref{ginc})
for the electron Green's function
in the presence of correlation.

From simple scaling relation it is straightforward to
recognize that
\begin{equation}
\Pi_{\rm c-c}
= Z\Pi(Z=0)=-ZN_0.
\end{equation}

The analytical expressions for $\Pi_{\rm c-i}$, $\Pi_{\rm i-i}$
are straightforward but more cumbersome since they involved
the explicit integration over the upper/lower Hubbard bands.
One obtains:
\begin{eqnarray}
\nonumber
\Pi_{\rm c-i}&=&-2N(0)^2(1-Z)\left[\left(ZE/2+U/2+(1-\frac{n}{2})E/2\right)
\right.
\times
\\
\nonumber
&\times&
\ln\left(ZE/2+U/2+(1-\frac{n}{2})E/2\right) +
\nonumber
\\
&-&\left(ZE/2+U/2-(1-\frac{n}{2})E/2\right) \times
\nonumber
\\
&\times&\ln\left(ZE/2+U/2-(1-\frac{n}{2})E/2\right) +
\nonumber
\\
&-&\left(U/2+\mu+(1-\frac{n}{2})E/2\right) \times
\nonumber
\\
&\times&
\ln\left(U/2+\mu+(1-\frac{n}{2})E/2\right) 
\nonumber
\\
&+&\left(U/2+\mu-(1-\frac{n}{2})E/2\right) \times
\nonumber
\\
&\times&
\ln\left(U/2+\mu-(1-\frac{n}{2})E/2\right)+ 
\nonumber
\\
&+&\left(U/2+nE/4+ZE/2\right) \times
\nonumber
\\
&\times&
\ln\left(U/2+nE/4+ZE/2\right)+
\nonumber
\\
&-&\left(U/2-nE/4+ZE/2\right) \times
\nonumber
\\
&\times&
\ln\left(U/2-nE/4+ZE/2\right)+
\nonumber
\\
&-&\left(U/2+nE/4-\mu\right) \times
\nonumber
\\
&\times&
\ln\left(U/2-nE/4-\mu \right)+
\nonumber
\\
&+&\left(U/2-nE/4+\mu\right) \times
\nonumber
\\
&\times&
\ln\left(U/2-nE/4+\mu\right)
\end{eqnarray}

The incoherent-incoherent contribution gives:
\begin{eqnarray}
\nonumber
\Pi_{\rm i-i}&=&-2N_0^2(1-Z)^2\left[(U+E/2)\ln(U+E/2)+
\right.
\\
\nonumber
&+&(U-E/2)\ln(U-E/2) +
\\
\nonumber
&-&(U+(n-1)E/2)\ln(U+(n-1)E/2) +
\\
\nonumber
&-&\left.(U+(1-n)E/2)\ln(U+(1-n)E/2)\right]
\end{eqnarray}

\subsection{Superconducting properties}

Here we report the explicit expressions for the coherent
contribution to different quantities in Eqs. (\ref{z})-(\ref{gap}).

Let us consider for instance $\eta_m[G_{\rm coh}]$.
In this case $\eta_m[G_{\rm coh}] = \eta_m^\Delta[G_{\rm coh}]$ and:
\begin{eqnarray}
\label{aeta}
\nonumber
\eta_m[G_{\rm coh}]&=&\int d\epsilon N(\epsilon) 
\frac{Z}{i\omega_m- Z \epsilon_{\bf k}-\mu}
 \frac{Z}{-i\omega_m - Z\epsilon_{\bf -k}-\mu}
\\
&=&\frac{Z}{E\omega_m}
\left[\arctan\left(\frac{ZE-\mu}{2\omega_m}\right)
+\arctan\left(\frac{ZE+\mu}{2\omega_m}\right)\right].
\end{eqnarray}

The expression (\ref{aeta}) corresponds just to the
$\eta_m(E)$ for an uncorrelated system with reduced spectral
weight $Z$ and rescaled bandwidth $ZE$:
$\eta_m[G_{\rm coh}](E) = Z \eta_m(ZE)$.
Similar considerations apply for the vertex and cross function:
$P[G_{\rm coh}](E,Q_c;n,m) = Z P(ZE,Q_c;n,m)$,
$C[G_{\rm coh}](E,Q_c;n,m) = Z C(ZE,Q_c;n,m)$,
where $P(E,Q_c;n,m)$ and $P(E,Q_c;n,m)$ in the absence
of electronic correlation were computed in Refs. \onlinecite{GPS,PSG,sgp}:

\begin{eqnarray}
\nonumber
&&P(E,Q_c;n,m)=T\sum_l D(\omega_n-\omega_l)\Bigg\{B(n,m,l)+
\\
\nonumber
&&
+ \frac{A(n,m,l)-B(n,m,l)(\omega_l - \omega_{l-n+m})^2}
{EQ^2_c} \times
\\
\nonumber
&&
\times \left[\sqrt{1+\left(\frac{2EQ^2_c}{\omega_l-\omega_{l-n+m}}\right)}
- 1 +
\right.
\\
&&
\left.
- \ln\left(\frac{1}{2}\sqrt{1+\left(\frac{2EQ^2_c}
{\omega_l - \omega_{l-n+m}} \right)^2}\right)\right]\Bigg\},
\end{eqnarray}
\begin{eqnarray}
\nonumber
&&C(E,Q_c;n,m)=D(\omega_n-\omega_l)D(\omega_l-\omega_m)\times
\\
\nonumber
&&\times
\Bigg\{2B(n,-m,l) +\arctan\left(\frac{2EQ^2_c}{|\omega_l-\omega_{l-n+m}|}\right)\times
\\
&&\times \frac{A(n,-m,l)-B(n,-m,l)(\omega_l-\omega_{l-n-m})^2}{EQ^2_c|\omega_l - \omega_{l-n-m}|}\Bigg\},
\end{eqnarray}
where
\begin{eqnarray}
\nonumber
A(n,m,l)&=&(\omega_l-\omega_{l-n+m})\left[
\arctan\left(\frac{E}{2\omega_l}\right)+\right.
\\
&-&
\left.
\arctan\left(\frac{E}{2\omega_{l-n+m}}\right)\right],
\\
\nonumber
B(n,m,l)&=&(\omega_l-\omega_{l-n+m})
\frac{E\omega_{l-n+m}}{2\left[(E/2)^2+\omega^2_{l-n+m}\right]^2}+
\\
&-&
\frac{E}{2[(E/2)^2+\omega^2_{l-n+m}]}.
\end{eqnarray}

\end{appendix}

\end{document}